\documentclass{sigplanconf}

\usepackage{amsmath}
\usepackage{graphicx}
\usepackage{listings}
\usepackage[htt]{hyphenat}

\begin{document}

\setlength{\pdfpageheight}{\paperheight}
\setlength{\pdfpagewidth}{\paperwidth}
\lefthyphenmin4
\righthyphenmin3

\conferenceinfo{PPPJ '14}{September 23-26, 2014, Cracow , Poland}
\copyrightyear{2014}
\copyrightdata{978-1-nnnn-nnnn-n/yy/mm}
\doi{nnnnnnn.nnnnnnn}

% Uncomment one of the following two, if you are not going for the
% traditional copyright transfer agreement.

%\exclusivelicense                % ACM gets exclusive license to publish,
                                  % you retain copyright

%\permissiontopublish             % ACM gets nonexclusive license to publish
                                  % (paid open-access papers,
                                  % short abstracts)

\titlebanner{banner above paper title}        % These are ignored unless
\preprintfooter{short description of paper}   % 'preprint' option specified.

\lstset{breaklines=true,basicstyle=\footnotesize\ttfamily} % Font style for Coded text.

%================= TITLE =================
\title{Design and Implementation of Parallel Debugger and Profiler for MPJ Express}

%================= Authors Area Start =================
\authorinfo{Aleem Akhtar}
           {National University of Sciences and Technology, Pakistan}
           {aleem.akhtar@seecs.edu.pk}
\authorinfo{Aamir Shafi}
           {National University of Sciences and Technology, Pakistan}
           {aamir.shafi@seecs.edu.pk}
\authorinfo{Mohsan Jameel}
           {National University of Sciences and Technology, Pakistan}
           {mohsan.jameel@seecs.edu.pk}
%================= Authors Area End =================

\maketitle

%================= Abstract =================%
\begin{abstract}

MPJ Express is a messaging system that allows computational scientists to write and execute parallel Java applications on High Performance Computing (HPC) hardware. Despite its successful adoption in the Java HPC community, the MPJ Express software currently does not provide any support for debugging and profiling parallel applications and hence forces its users to rely on manual and tedious debugging/profiling methods. Support for such tools is essential to help application developers increase their overall productivity. To address this we have developed debugging and profiling tools for MPJ Express, which are the main topic of this paper. Key design goals for these tools include: 1) maintain compatibility with existing logging, debugging, and visualizing tools, 2) build these tools by extending existing debugging/profiling tools instead of reinventing the wheel. The first tool, named {\em MPJDebug}, builds on the open-source Eclipse Integrated Development Environment (IDE). It provides an Eclipse-based plugin developed using the Eclipse Plugin Development Environment (PDE). The default Eclipse debugger currently does not support debugging parallel applications running on a compute cluster. The second tool, named {\em MPJProf}, is a utility based on Tuning and Analysis Utility (TAU)---an open-source performance evaluation tool. Our goal here is to exploit TAU to profile Java applications parallelized using MPJ Express by generating profiles and traces, which can later be visualized using existing tools like {\tt paraprof} and {\tt Jumpshot}. Towards the end of the paper, we quantify the overhead of using MPJProf, which we found to be negligible in the profiling stage of parallel application development.

\end{abstract}
%================= Abstract End =================

%================= ACM Categories =================

% general terms are not compulsory anymore,
% you may leave them out


%================= ACM Categories End =================

%================= Introduction Section Start =================
\section{Introduction}

The Message Passing Interface (MPI) standard \cite{mpispec} has become the \emph{de facto} API for programming High Performance Computing (HPC) hardware including commodity clusters. The current version of the MPI standard supports traditional programming languages like C and Fortran by providing bindings for these languages. The two most popular implementations of MPI include MPICH \cite{mpich} and Open MPI \cite{openmpi}. On the other hand, modern languages with features of object orientation, modularity, maintainability and portability have been treated with cynicism, mostly because of their poor computing performance and lack of high performance communication support \cite{Blount1998}. This criticism is not justified anymore because most modern languages and their compilers and runtime environments have witnessed manifold performance improvements. An example of one such modern programming language is Java. By the use of Just-in-Time (JIT) compilers the performance gap between Java byte code and native code is becoming negligible \cite{Taboada2009}. The emergence of many popular and successful Java messaging libraries like mpiJava \cite{carpenter}, FastMPJ \cite{Taboada2012} and MPJ Express \cite{Shafi2009} have successfully helped decrease communication gap between C/Fortran and Java applications on HPC hardware.

MPJ Express is an MPI-like---implements the mpiJava 1.2 API---messaging library with an active user community. The software is capable of executing in two modes named {\em cluster} and {\em multicore} modes. In the cluster mode, parallel applications execute in a typical cluster environment where multiple processing elements communicate with one another using a fast interconnect like Gigabit Ethernet or other proprietary networks like Myrinet and InfiniBand. In the multicore mode, the parallel Java application executes on a single system comprising of shared memory or multicore processors.

Despite its successful adoption in the Java HPC community, the MPJ Express software currently does not provide any support for debugging and profiling parallel applications and hence forces its users to rely on manual and tedious debugging/profiling methods, which require manually adding printing/logging/timing statements and constant recompilation of end user application. Support for such tools is essential to help application developers increase their overall productivity. In addition, manually debugging/profiling parallel applications is a complex, and challenging undertaking due to multitude of challenges including large scale parallelism, non-determinism, communication delays, synchronization requirements, concurrency control, and process locality.

To address this, we have developed debugging and profiling tools for MPJ Express named {\em MPJDebug} and {\em MPJProf} respectively. Key design goals for these tools include: 1) maintain compatibility with existing logging, debugging, and visualizing tools, 2) build debugging/profiling tools by extending existing debugging/profiling tools instead of reinventing the wheel. The first tool, named MPJDebug, builds on the open-source Eclipse Integrated Development Environment (IDE). MPJDebug is an Eclipse-based plugin---developed using the Eclipse Plugin Development Environment (PDE)---that allows MPJ Express users to execute and debug parallel Java applications running in the multicore mode or cluster mode. The default Eclipse debugger currently does not support debugging parallel applications running on a compute cluster. This is attractive because MPJ Express users can now utilize standard debugging features-including stepping, conditional and exception breakpoints, watch points and drop to frame---for their parallel Java applications. The second tool, named MPJProf, is a utility based on Tuning and Analysis Utility (TAU)---an open-source performance eval-uation tool. Our main goal is to provide MPJ Express users with a tool to analyze performance of their parallel Java applications. We achieve this by exploiting TAU to generate profiles and traces, which can later be visualized using existing tools like {\tt paraprof}, {\tt Jumpshot}, and {\tt pprof}. Since we are building on a popular existing tool, different features and views are available for end users that include 3D visualization, threads-based and functions-based display.

Towards the end of the paper, we quantify the overhead of our profiling tool by employing variety of performance tests including basic latency and bandwidth benchmarks for point-to-point communication and Java NAS parallel benchmarks (NPB). Our results indicate that the MPJProf tool only adds a negligible overheads, which is due to generation of profiling information by parallel processes.

Rest of the paper is organized as follows. Section 2, discusses related work. Section 3 and 4 present implementation details of MPJDebug and MPJProf tools, respectively. This is followed by evaluating performance of the MPJ Express software with MPJProf in Section 5. Finally Section 6 concludes and discusses future work.

%================= Introduction Section END =================

%================= Related Work Section Start =================
\section{Related Work}
This section provides an overview of existing parallel debugging and profiling tools and also motivates the need for these tools in the context of our MPJ Express software. We begin our discussion with a review of debugging tools followed by profiling tools.

\subsection{Debugging Tools}

\emph{TotalView} \cite{Totalview} is a powerful tool used for debugging parallel programs running on UNIX, Linux and Mac OS X. It supports the usual HPC application languages including C, C++ and Fortran. \emph{Allinea Distributed Debugging Tool (DDT)} \cite{ddt} is a commercial software that is capable of debugging scalar, multithreaded and large-scale parallel applications. The tool supports C, C++, Fortran, Coarray Fortran, UPC, CUDA and OpenMP. \emph{Eclipse} \cite{Pawel2003}---an open-source IDE---features a built-in Java debugger that provides standard debugging features like breakpoint setting, step execution, suspend/resume threads and variable inspection. It is also capable of debugging remote applications. Eclipse IDE and debugger also support development/debugging in other popular languages including C, C++, and Python.

TotalView and DDT are the two most popular debugging tools for debugging C/C++/Fortran parallel MPI applications. Since these tools do not yet support Java, they cannot be used for debugging Java MPI applications parallelized with MPJ Express or FastMPJ. Being commercial tools, it is not possible for us to extend these tools. On the other hand, Eclipse being an open-source high-quality IDE is a good candidate to debug parallel Java MPI applications running in cluster and multicore modes. Our investigations suggest that it is straightforward to debug Java MPI applications running in the multicore mode. On other hand, the vanilla Eclipse debugger does not support debugging parallel Java applications running in the cluster mode---we add this feature in Eclipse as part of our main contribution in this paper.

\subsection{Profiling Tools}

\emph{JProfiler} \cite{JProfiler} is a commercial tool for profiling Java SE/SE applications. Salient features of the tool includes an intuitive Graphical User Interface (GUI) that allows users to find performance bottlenecks, pin down memory leaks, and resolve threading issues. JProfiler is capable of profiling a single Java Virtual Machine (JVM) application. However, an MPJ Express application typically consists of a group of Java processes---executing in separate JVMs. Another similar profiling suite is the \emph{JProbe} \cite{JProbe} software, which contains various tools to analyze performance of Java applications. These tools allow end-users to conduct post-mortem analysis of applications. JProbe is also a commercial product and it is thus not possible to extend it to support parallel MPJ Express programs.

\emph{Tuning and Analysis Utility} (TAU) \cite{TAU} is a portable profiling and tracing toolkit for performance analysis of parallel programs written in Fortran, C/C++, UPC, Java and Python. TAU provides graphical and command line tools such as {\tt paraprof} and {\tt pprof} to visualize profiling results in nodes/threads and aggregated format. Typically users utilize open-source and free tools like {\tt Vampir}, {\tt Jumpshot}, and {\tt Paraver} to visualize event traces. TAU also support some well-known parallel programming implementations like MPICH, Open MPI and mpiJava. As part of this paper, one of the objectives is to exploit TAU to profile Java applications parallelized using MPJ Express.
%================= Related Work Section END =================

%================= Implementation of MPJDebugSection Start =================
\section{Implementation of MPJDebug}

This section details the implementation of our debugging tool for MPJ Express called MPJDebug. The first sub-section presents requirements from the perspective of our users. This is followed by an overview of the Eclipse plugin development architecture. Finally, we layout implementation details of our plugin that allow us to {\em launch} or {\em debug} parallel Java applications.

\subsection{Requirements of Debugger}

In the context of MPJ Express users, the MPJDebug tool must support the following requirements:

\begin{enumerate}
  \item \emph{Basic features}: Provide basic debugging features such as stepping, breakpoints and suspend/resume.
  \item \emph{Ease to use}: Must be be usable---in terms of ease of use---by the application developers.
  \item \emph{Scalable}: Must be scalable enough to allow debugging large parallel programs executing on hundreds and thousands of processors.
  \item \emph{Remote debugging}: Must support MPJ Express multicore and cluster modes. When executing in cluster mode, a debugger is required to be able to connect to remote parallel processes running on cluster nodes.
\end{enumerate}

With these goals in mind, a plausible approach to develop a debugger for MPJ Express is to extend a sequential debugger, like provided by Eclipse IDE, for compute clusters \cite{Steve1994}.

\subsection{Eclipse Plugin Architecture}

\begin{figure}[htp]
  \center
  \includegraphics [width=0.41\textwidth]{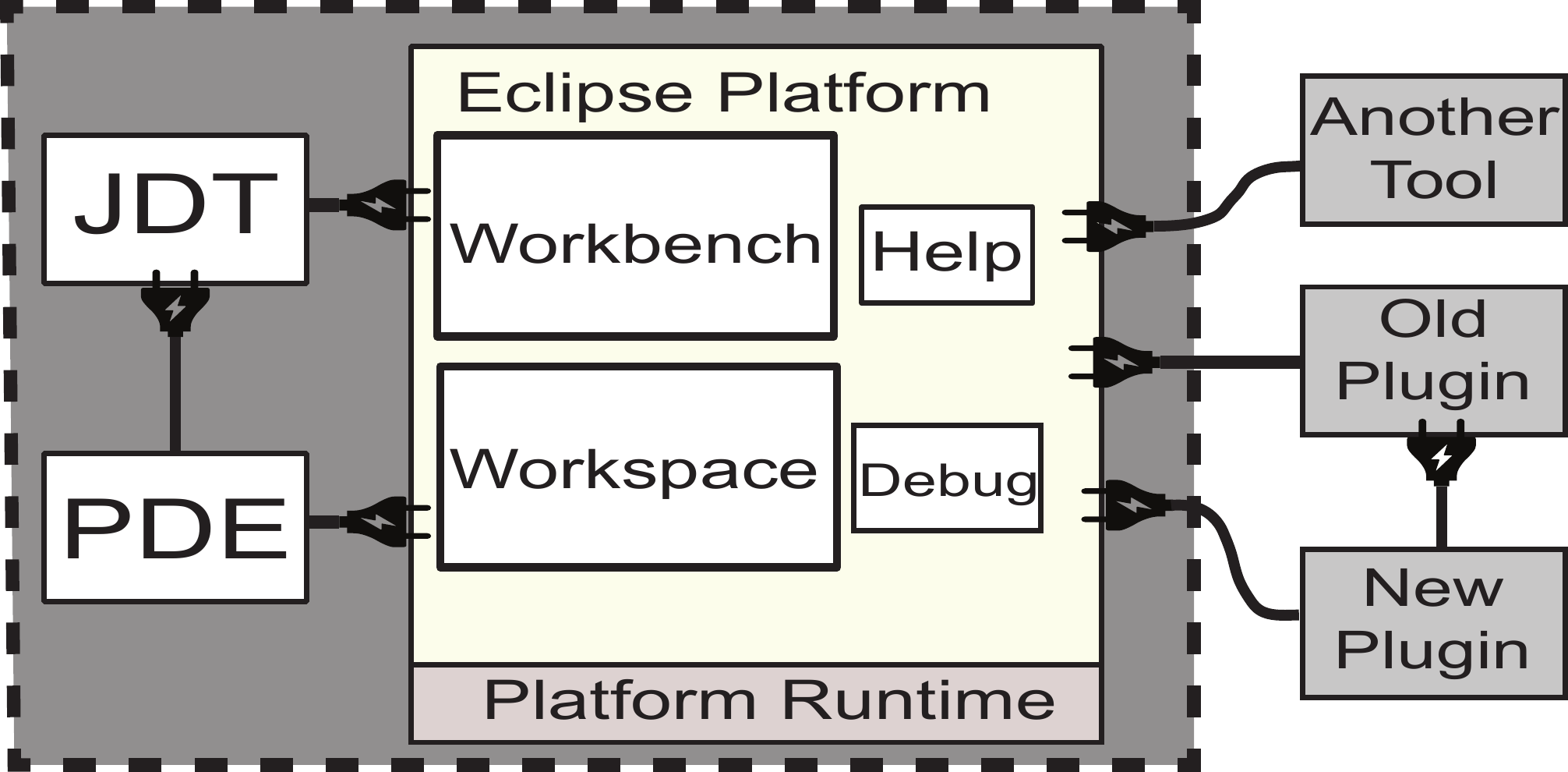}
  \caption{Eclipse Plugin Architecture}
  \label{eclipsePluginArchictecture}
\end{figure}

The Eclipse Platform is an open framework, which consists of core technologies like Java Development Tools (JDT) and Plugin Development Environment (PDE) as shown in Figure ~\ref{eclipsePluginArchictecture}. The core platform consists of some essential components including a platform runtime---again depicted in Figure ~\ref{eclipsePluginArchictecture}. The functionality of the core Eclipse platform can be extending by building new plugins using the Plugin Development Kit (PDK) alongwith JDT and PDE. Owing to its modular architecture, many essential development tools are provided by Eclipse community as plugins. These plugins interact with one another and the core Eclipse platform using standard and published interfaces called {\em Extension Points}---these are depicted by power sockets in Figure ~\ref{eclipsePluginArchictecture}.

We follow the same approach for building the MPJDebug tool. Each plugin is developed as a self-contained software module, which contains the plugin manifest file---named {\tt plugin.xml}. This file is written in the XML format and is typically load first by the Eclipse platform to customize the new plugin. The manifest file contains all necessary configuration information including details of extension points and display items like icons/menu items.

\begin{figure}[htp]
  \center
  \includegraphics [width=0.47\textwidth]{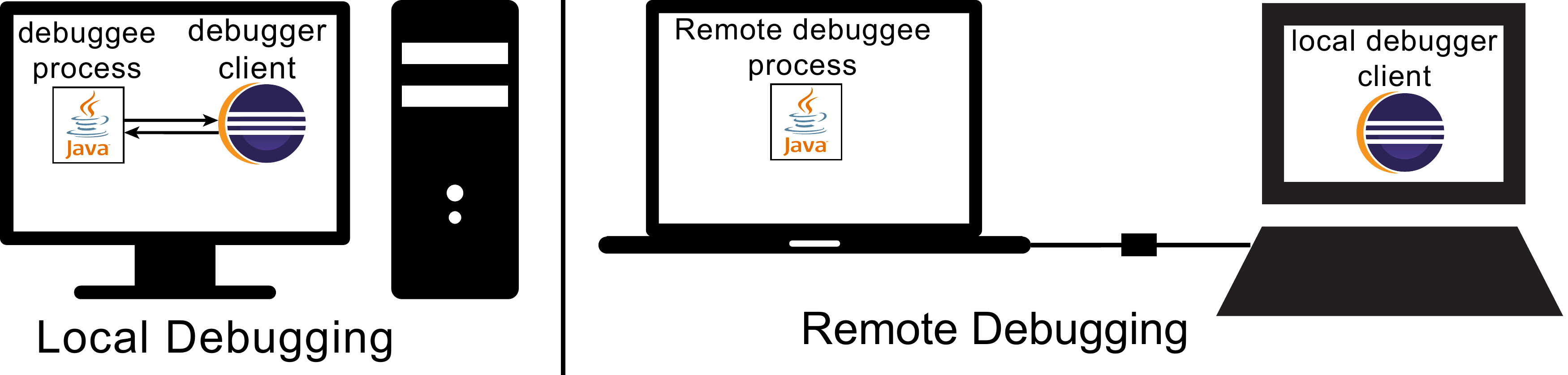}
  \caption{Local and Remote Debugging}
  \label{LocalRemoteDebugging}
\end{figure}

Eclipse workbench cannot launch or debug applications on its own. Launching means executing a program without it being suspended or examined and debugging means, execution may be suspended and resumed, variables may be inspected, and expressions may be evaluated. Eclipse workbench uses different set of plugins to acquire this feature. Debugger is one such plugin that Eclipse uses to launch or debug applications. Client/server design of debugger can debug programs running on local machines and running remotely on other systems in the network. The debug client runs on local workstation and debugger server runs on the same system as the program you want to debug.  This could be a program launched on local workstation (local debugging) or a program started on a computer that is accessible through network (remote debugging). Left side of Figure ~\ref{LocalRemoteDebugging} display local debugging where both debuggee process/program and debugger client are at the same machine and remote debugging is shown in the right side of ~\ref{LocalRemoteDebugging} where debuggee process/program is at remote machine while debugger client is running inside the workbench on local machine and both machines are connected through network. Eclipse has a special Debug view that displays stack frame for suspended threads for each target you are debugging.

\subsection{Implementation of launching feature}
 Implementation of MPJDebug is based on Eclipse plugin architecture. It is developed by extending various features of Eclipse debugger to MPJDebug. To gain launching feature, MPJDebug extends Eclipse debugger feature of launching . Implementation of launching feature in MPJDebug is breakdown in different steps. Each step is described below.

\textbf{Step-1: Extension of LaunchConfigurationType}:
 Eclipse debugger uses {\tt LaunchConfigurationTypes} extension point to launch Java applications. MPJDebug uses this extension point to achieve launching feature. This extension point provides a configurable mechanism for launching applications. Each launch configuration type has a name, supports one or more modes (run and/or debug), and specifies a delegate responsible for the implementation of launching an application \cite{launch}. In Eclipse debugger, we have ``Java Application" and ``Remote Java Application" configuration types. This extension is declared in plugin manifest file. Code snippet for that extension is provided below

\begin{lstlisting}
<extension point="org.eclipse.debug.core.launchConfigurationTypes">
    <launchConfigurationType
        name="MPJExpress Application"
        delegate="JavaLaunchDelegate"
        modes="run,debug"
        id="mpjExpress">
    </launchConfigurationType>
</extension>
\end{lstlisting}

\textbf{Step-2: Declaration of delegate}:
Delegate attribute of {\tt LaunchConfigurationType} is the most important one as it specifies fully qualified name of class that implements the interface {\tt ILaunchConfigurationDelegate}. Second step is to create delegate class that implement interface {\tt ILaunchConfigurationDelegate}. This class has a method {\tt launch} which takes all required information through parameters and launches one of the modes defined in configuration type \cite{Darin2003}. The {\tt launch} method is the first function that is invoked when an application is executed in new configuration.

\begin{lstlisting}
launch(ILaunchConfiguration configuration,
            String mode, ILaunch launch,
            IProgressMonitor monitor)
    throws CoreException
\end{lstlisting}

\textbf{Step-3: Definition of launch method}:
Launch method takes different parameters of which {\tt ILaunchConfiguration} parameter is important one as it contains all pertinent information that is related to launching of an application such as program arguments and VM arguments. In case of launching of MPJ Express application this configuration parameter contains information like number of processes, path to root directory of MPJ Express, device type and any additional parameters. All of this information is provided to configuration through MPJ parameters tab.

\begin{figure}[htp]
  \center
  \includegraphics [width=0.47\textwidth]{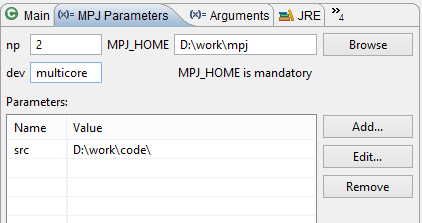}
  \caption{MPJ Parameters Tab}
  \label{mpjParameterTab}
\end{figure}

\textbf{Step-4: Implementation of MPJ parameters tab}:
This tab is implemented using extension {\tt launchConfigurationTabGroup}. This tab acts as graphical user interface where user can provide different options to launch parallel Java application. Options available in this tab are name of device, number of processes, path to MPJ Express root directory and support for different parameters. MPJ parameters tab is added as a part of {\tt launchConfigurationTabGroup} so options from other tabs such as program arguments and VM arguments are available to the users. Figure ~\ref{mpjParameterTab} depicts MPJ parameters tab is configured to launch parallel Java application in multicore mode MPJ Express.

\textbf{Step-5: Final Launching}:
After all pertinent information is provided in MPJ parameters tab and application is launched in run mode, {\tt launch} method of delegate class is invoked. In {\tt launch} method, information contained in configuration parameter is extracted and is modified to form a command that is used for launching MPJ Express applications.  That command is then appended to VM argument of configuration. Finally configuration is launched using {\tt IVMRunner}.
\subsection{Implementation of debugging feature}
{\tt LaunchConfigurationTypes} can execute Java applications in debug mode as well. To debug Java applications we need to enable debugger agent by passing following options in VM arguments of configuration.
\begin{lstlisting}
-agentlib:jdwp=transport=dt_socket,server=y,suspend=n,address=8000
\end{lstlisting}

Parameter {\tt address} acts as transport address for the socket connection between debugger client and debuggee process (from now on MPJ process). Java Debug Wire Protocol (JDWP) is used as communication protocol between MPJ process and debugger client. If {\tt server=y}, then MPJ process is launched in debug mode and listen on port (specified in transport address) for debugger client to connect to it. These debug options are shifted to runtime system of MPJ Express where transport address for each process is set. MPJDebug is capable of debugging parallel Java applications in following two modes.

\begin{figure}[htp]
  \center
  \includegraphics [width=0.47\textwidth]{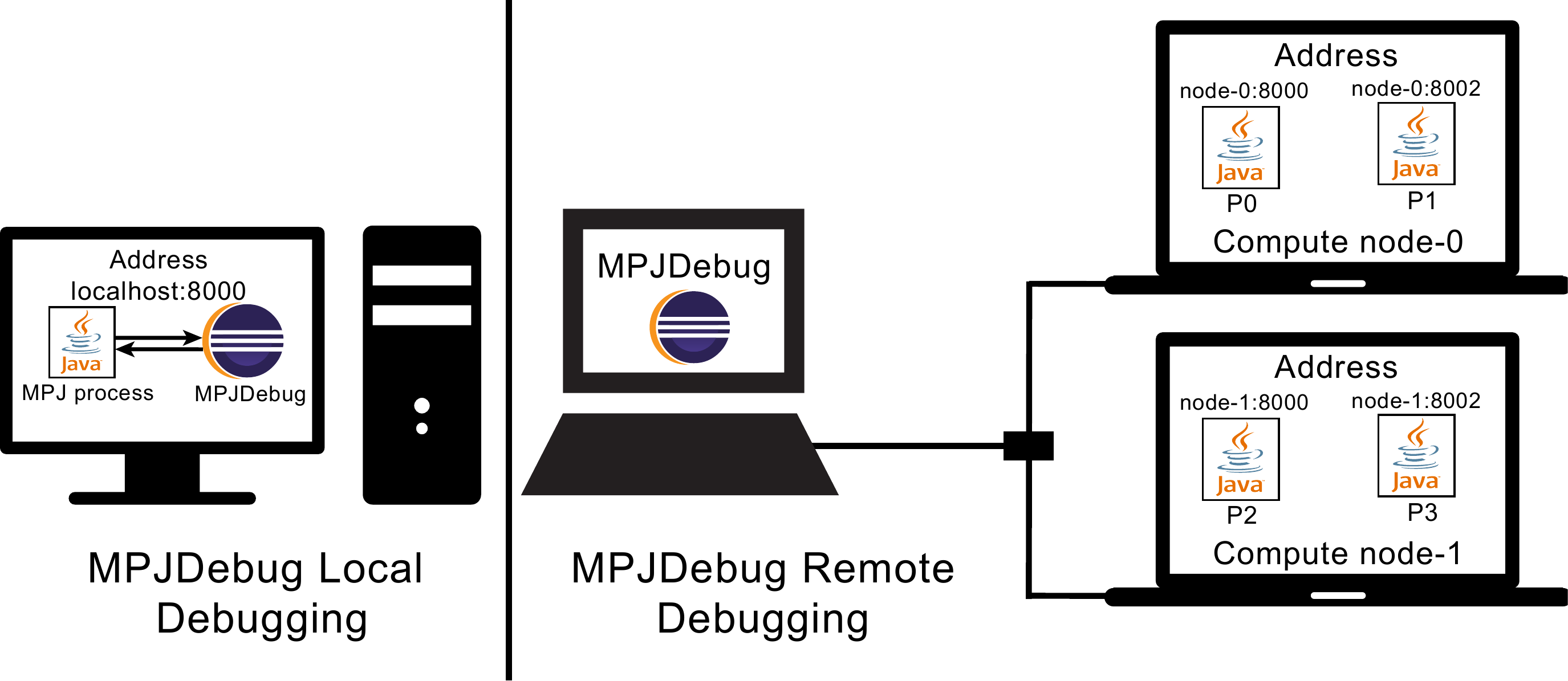}
  \caption{MPJDebug Remote Debugging}
  \label{remoteDebugging}
\end{figure}

\textbf{Local Debugging} means both MPJ process being debugged and debugger client are at the same workstation. Debugging of parallel Java applications running in multicore mode of MPJ Express come under this category. To launch parallel Java applications in debug mode we provide a debug parameter with port as value through MPJ parameters tab. Value of this parameter is passed to runtime system of MPJ Express where it is used as value for address in debug options. MPJ process starts listening on set port. Debugger then connects to the listening port and application is launched in debug mode where different features can be used to debug application. Local debugging by MPJDebug can be seen in left side of Figure ~\ref{remoteDebugging} where MPJ process and MPJDebug are at the same machine and port assigned to MPJ process is 8000.

%\begin{figure}[htp]
 % \center
 % \includegraphics [width=0.15\textwidth]{localDebugging.png}
 % \caption{MPJDebug Local Debugging}
 % \label{localDebugging}
%\end{figure}

\textbf{Remote Debugging} means MPJ process being debugged is at some remote machine (accessible through network) and debugger client is at local workstation. Debugging of parallel Java application running in cluster mode of MPJ Express come under this category. MPJ processes are distributed across the nodes of cluster, and there is possibility of more than one MPJ process will launch at one node. So a different port for each MPJ process launched at one node of cluster is required otherwise debugger throws an error stating ``address already in use". Same port is used by MPJ processes launched at different nodes. To make sure different port is assigned to each MPJ processes at one node, we use the value provided against debug parameter in MPJ parameters tab and set a different address for each MPJ process. Formula used to generate new address is {\tt (initial\_debug\_port + (2 * n))} where `n' ranges from 0 to maximum number of MPJ processes that are to be launched at one node. If user provides {\tt initial\_debug\_port}  as 8000 and number of nodes are 2 then each node will host 2 processes and following ports will be set for each process. Right side of Figure ~\ref{remoteDebugging} is an illustration of this example.

\begin{lstlisting}
Node-0 Process-0:8000; Node-1 Process-2:8000
Node-0 Process-1:8002; Node-1 Process-3:8002
\end{lstlisting}

 At the same time we are writing these ports along with names or IP addresses of nodes in a configuration file called {\tt mpjdev.conf}. This file is accessible to each compute node and MPJDebug. It contains information of each compute node including IP address, process rank and debug port. MPJ Express runtime system use this file to assign debug port to each process. Once port assigning is completed, MPJ processes are launched and start listening on respective port.

MPJDebug reads {\tt mpjdev.conf} file and retrieve IP addresses and their respective ports. Finally Java VM Connector is used to establish a connection to remote processes and JDWP communicate with MPJ processes using their respective port. Application can be further debugged using different debugging features.

%================= Implementation of MPJDebug Section End =================

%================= Implementation of MPJProf Section Start =================
\section{Implementation of MPJProf}
This section details the implementation of our profiling tool for MPJ Express called MPJProf. The first sub-section presents requirements from the perspective of our users. This is followed by an overview of TAU. Finally, we layout implementation details of our plugin that allow us to {\em profile} parallel Java applications.

\subsection{Requirements of Profiler}
In the context of MPJ Express users, the MPJProf tool must support the following requirements:

\begin{enumerate}
  \item \emph{Basic features}: Provide basic profiling features such as time analysis of methods and statements
  \item \emph{Instrumentation}: Must support both automatic and manual instrumentation
  \item \emph{Scalable}: Must be scalable enough to allow profiling of large parallel programs executing on hundreds and thousands of processors.
  \item \emph{MPJ Express modes}: Must support profiling of MPJ Express multicore and cluster modes.
  \item \emph{Tracing}: Must be capable of generating event traces of parallel java applications.
\end{enumerate}

With these goals in mind, a plausible approach to develop a profiler for MPJ Express is to exploit features of open-source performance analysis tool like TAU.

\subsection{Tuning and Analysis Utility (TAU)}
Tuning and Analysis Utility (TAU) is a portable profiling and tracing toolkit for performance analysis of parallel programs written in Fortran, C, C++, UPC, Java, Python. TAU can profile parallel Java applications by collecting performance data of each method, statements and basic blocks for each thread, context, and node in use by an application. Using performance data, TAU can generate profiles for users which contain wealth of performance information. These profiles can give information about inclusive and exclusive time spent in each function in different time units, number of times function was called or how many functions were invoked by each function. Using this information user can easily identify performance bottlenecks in their applications.

Profiles generated by TAU, follow a special naming scheme {\tt profile.<node>.<context>.<thread>}. TAU provides graphical and command line tools such as {\tt paraprof} and {\tt pprof} to visualize profiling results in nodes/threads and aggregated format. TAU can also generate event traces which follow the same name scheme {\tt trace.<node>.<context>.<thread>}. Traces display when an event took place along a timeline. Typically users utilize open-source and free tools like {\tt Vampir}, {\tt Jumpshot}, and {\tt Paraver} to visualize event traces. TAU also support some well-known parallel programming implementations like MPICH, Open MPI and mpiJava. As part of this paper, one of the objectives is to exploit TAU to profile Java applications parallelized using MPJ Express.

\subsection{MPJProf in Multicore Mode}
Profiling of parallel Java applications running in multicore mode using TAU is straightforward. TAU uses {\tt tau\_java} to generate profiles for multithreaded Java applications. {\tt tau\_java} instruments Java applications at runtime using Java Virtual Machine Tool Interface (JVMTI). In multicore mode we have one node and multiple threads so generated profiles will be like; {\tt profile.0.0.t} where {\tt t} ranging from 0 to maximum number of threads being launched at the node. Use of {\tt tau\_java} is implemented in runtime system of MPJ Express software. If user has selected to profile parallel Java application then MPJ Express runtime system enable {\tt tau\_java} to start profiling.
\subsection{MPJProf in Cluster Mode}
TAU can generate profiles for applications running in multicore mode by using {\tt tau\_java} but it cannot profile parallel Java applications running in cluster mode of MPJ Express. There is support for mpiJava in TAU \cite{Shende2001} to perform analysis of applications running at compute cluster. But mpiJava is implemented as a set of JNI wrappers to native MPI packages where MPJ Express is pure Java implementation of MPI. Profiles are not generated as per expectation with MPJ Express because value for ``node" is hard-coded to zero in TAU source code and there is no option to change this value at runtime. Node represents process and in case of cluster mode, processes are distributed across nodes of cluster. So we need to change value of node according to processes. To achieve that we added {\tt \textbf{-tau:node<NodeID>}} configuration option in {\tt tau\_java}. {\tt NodeId} represents rank of MPJ process. When TAU runtime encounters this option it changes default value of node to option value. This setting of value is done in {\tt tauJVMTI.cpp}
\begin{lstlisting}
if option is node {
	set tau profile node to option value
}
\end{lstlisting}
This will change profile node to node id and profiles for that node will be generated.
\footnote{This change was proposed to TAU developers. They approved it and added it in their beta release version 2.23.1b of TAU.}

In MPJ Express runtime if user has provided option to profile parallel Java application we start setting of profile node values by passing rank of each process to TAU runtime. As an example, four processes are distributed across two machines, so generatd profiles will be like {\tt profiles.n.0.t}, where {\tt t} ranges from 0 to total number of processes and {\tt t} ranges from 0 to maximum threads launched by each node. Once profiles are generated, these can be viewed using graphical interface {\tt paraprof} or command line interface {\tt pprof}.

%================= Implementation of MPJProf Section End =================

%================= Implementation of Performance Analysis Section Start =================
\section{Performance Analysis for NPB}
MPJProf supports both profiling and tracing performance analysis for MPJ Epxress applications. Performance analysis results give user statistics of performance metrics and performance behavior. We performed analysis for Java NAS parallel benchmarks (NPB) ~\cite{Mallon2009} to provide useability of MPJProf. We used NPB kernel {\tt IS} with workload {\tt A} and run it under four processes. However, it should be understood MPJProf can be extended to larger number of processes.

\begin{figure}[htp]
  \center
  \includegraphics [width=0.47\textwidth]{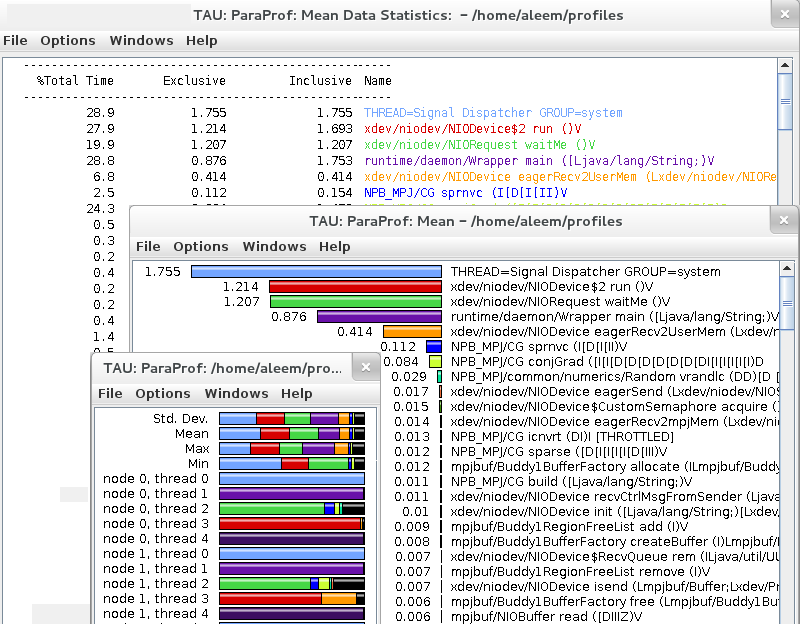}
  \caption{Various output windows of TAU's Paraprof}
  \label{profilingWindows}
\end{figure}

In Figure ~\ref{profilingWindows}, we see various output windows of TAU's profile browser {\tt paraprof}. For each profiling window metric is time and units are in seconds. We can see mean statistics of exclusive and inclusive time for all threads in Mean Data Statistics window sorted by exclusive time. Similarly we can view exclusive time for each node and thread in bar chart windows. Using this profiling data user can also see threads performing MPJ Express module and background JVM tasks which directly are not possible.

%================= Implementation of Performance Analysis Section End =================
%================= Implementation of Evaluation Section Start =================
\section{Evaluation}
 In this section, we quantify the overhead of our profiling tool by employing variety of performance tests including basic latency and bandwidth benchmarks for point-to-point communication and Java NAS parallel benchmarks (NPB). We performed evaluation on following two test environments. The first test environment (from now on RCMS) consisted of a 32 node cluster hosted at RCMS-NUST, Pakistan. Each compute node contains two quad-core Intel Xeon E5520 processors with a main memory of 24G Bytes. The nodes are connected via Gigabit Ethernet. Its software environment consisted of the Oracle JDK 1.7.0\_25 and TAU 2.23.1b. The second test environment (from now on 1GE) consisted of four machines, each having Intel\textsuperscript{\textregistered}  Core\texttrademark  i5-3470 CPU with 3.20GHz and 8G Byte of memory. All four machines are connected through 1G Ethernet connection. Software environment of these machines consist of Oracle JDK 1.7.0.\_25 and TAU 2.23.1b. All systems are configured for optimized performance.

\begin{figure}[htp]
  \center
  \includegraphics [width=0.48\textwidth]{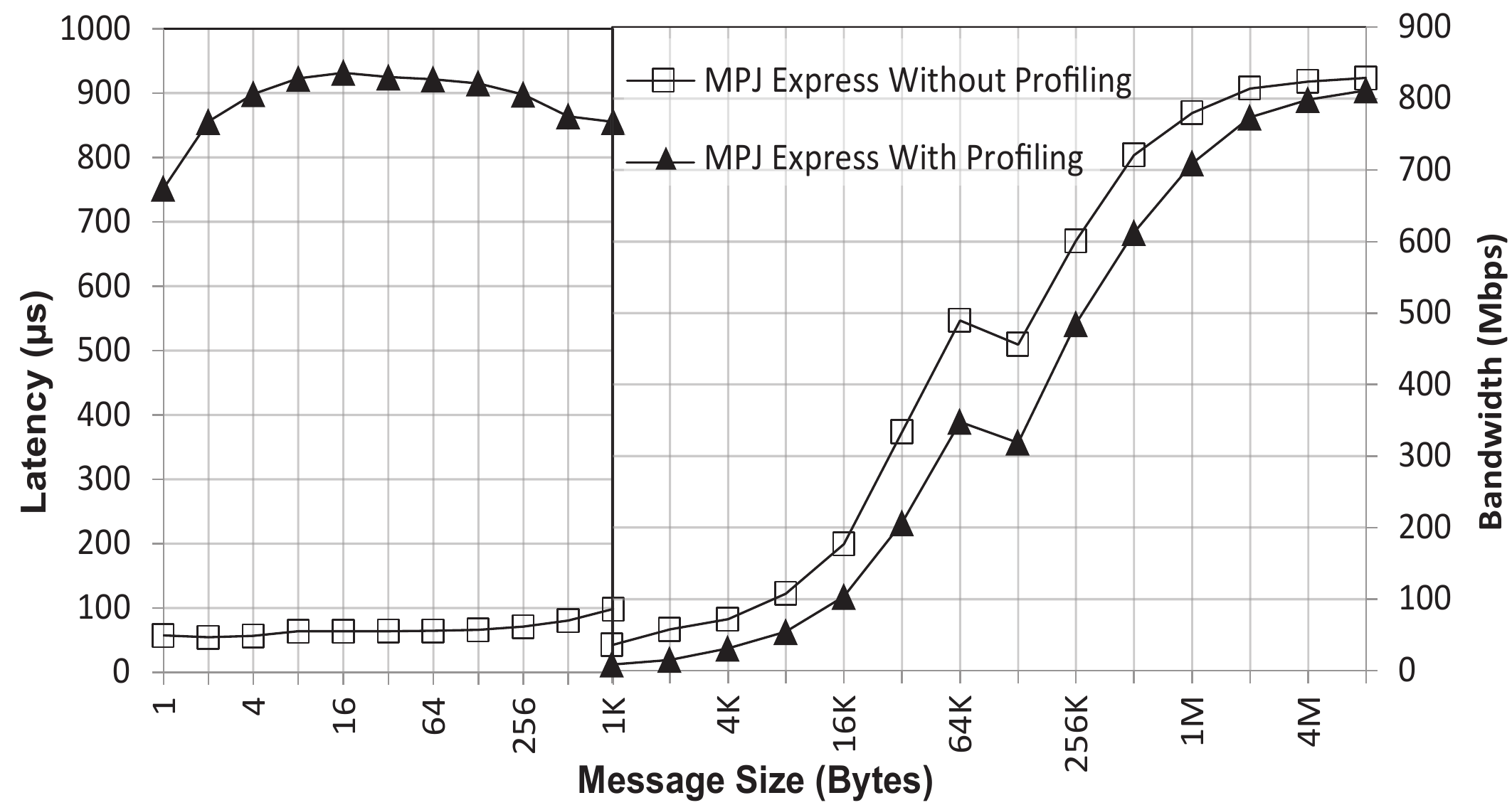}
  \caption{Latency and Bandwidth over Gigabit Ethernet}
  \label{LatencyBandwidth}
\end{figure}

We performed standard latency and bandwidth test on RCMS. Right side of Figure ~\ref{LatencyBandwidth} show throughput (bandwidth in Mbps) comparison across Gigabit Ethernet. MPJ Express achieves 83\% of maximum bandwidth when executed without MPJProf. There is performance loss when bandwidth test is run using MPJProf. Maximum bandwidth achieved in that case is 81\%. Left side of Figure ~\ref{LatencyBandwidth} shows the latency (transfer time for one byte in ${\mu}$s) comparison across Gigabit Ethernet. The latency for MPJ Express without profiling is 57.3${\mu}$s and with profiling is 752${\mu}$s. The reason for higher latency with MPJProf is due to generation of profiling information by parallel processes.

\begin{figure}[htp]
  \center
  \includegraphics [width=0.47\textwidth]{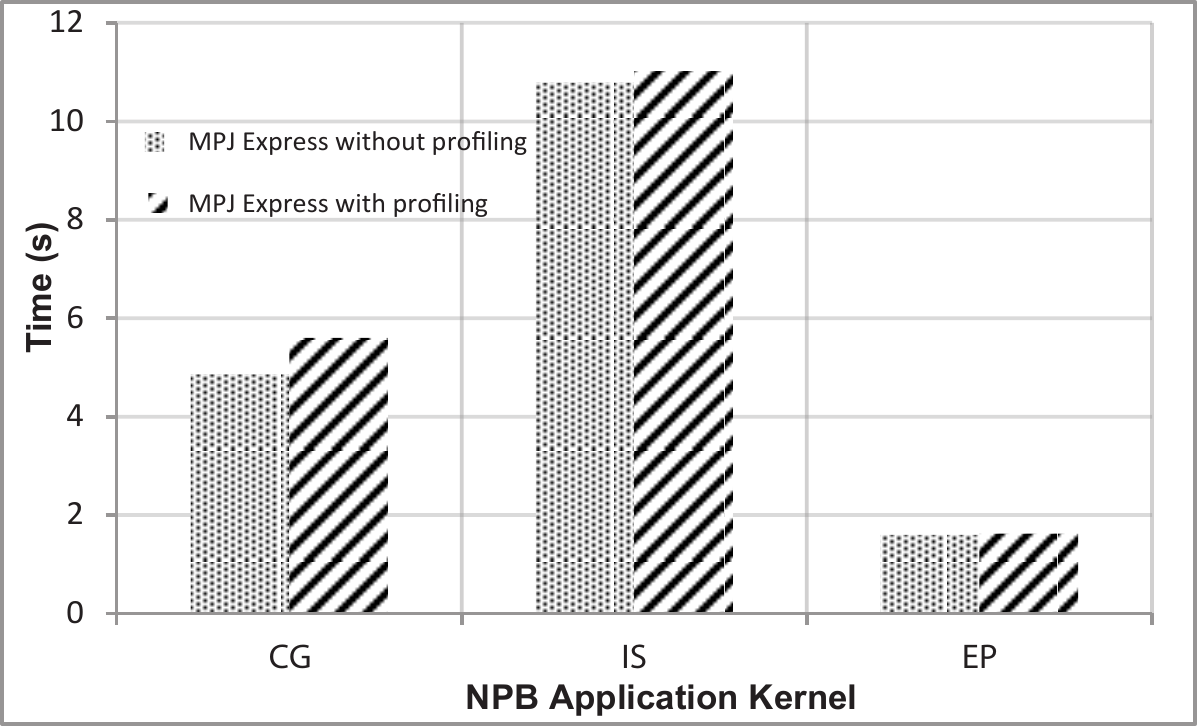}
  \caption{Overhead comparison for NPB }
  \label{NPBOverhead}
\end{figure}

We evaluated the performance of Java NAS parallel benchmarks (NPB) kernels on 1GE. We chose three NPB kernels namely {\tt CG}, {\tt IS} and {\tt EP} and ran test on workload Class A on total of 16 processes using MPJ Express. Figure ~\ref{NPBOverhead} shows performance analysis results time (in seconds) against NPB kernels using with and without MPJProf. We observed 15\% of overhead added in kernel {\tt CG}, 2\% of overhead added in kernel {\tt IS} and 1.5\% of overhead added in kernel {\tt EP}. Our results indicate that the MPJProf tool only adds a negligible overheads, which is due to generation of profiling information by parallel processes.

%================= Implementation of Evaluation Section End =================

%================= Implementation of Conclusion and Future Work Section Start =================
\section{Conclusions}

MPJ Express is a Java messaging system that allows parallelizing applications on distributed memory platforms including compute clusters. In this paper we presented debugging and profiling tools named MPJDebug and MPJProf respectively. The main goal is to increase overall productivity of MPJ Express application developers. The first tool, MPJDebug, builds on the open-source Eclipse IDE that allows extending functionality of the core platform by building Eclipse plugins. MPJDebug has been developed as an Eclipse plugin and supports debugging parallel applications executing in multicore and cluster modes. The second tool, MPJProf, is a utility based on Tuning and Analysis Utility (TAU)---an open-source performance evaluation tool. MPJProf exploits TAU to profile Java applications parallelized using MPJ Express by generating profiles and traces, which can later be visualized using existing tools like {\tt paraprof} and {\tt Jumpshot}. We quantified the overhead of using MPJProf, which we found to be negligible in the profiling stage of parallel application development.
%Current versions of MPJ Express Debugger and Profiler have no support for native device ~\cite{Bibrak14} implemented for MPJ %Express software. Next versions of these tools will be updated to work with native device. Users of MPJ Express also use Netbeans %IDE for development of parallel Java applications, so there is need of debugging tool/plug-in for Netbeans IDE.

%================= Implementation of Conclusion and Future Work Section End =================

%================= Implementation of MPJProf Section Start =================

% We recommend abbrvnat bibliography style.

%\input{PPPJ_Submission.bbl}
%\bibliographystyle{unsrt}

% The bibliography should be embedded for final submission.

%\bibliography{PPPJ_Submission}

\end{document}